\documentclass[12pt]{article}    %-*-latex-*-
\newtheorem{thm}{Theorem}[section]       % Numbered within each section
\newtheorem{proposition}[thm]{Proposition}
\newtheorem{corollary}[thm]{Corollary}
\newtheorem{definition}[thm]{Definition}
\newtheorem{remark}[thm]{Remark}          % Numbered along with thm
\def\dir#1.{x^{N-{#1}}y^{#1}}
\def\dirs#1.{x^{N-{(#1)}}y^{#1}}  % dirac symmetric bracket
\def\FFF#1#2{K^{(#1)}_{#2}}
\def\tr{\mathrm{tr}\,}

\def\eop{\vrule height 5pt width 5pt depth .5pt}%(JUST THE BLACK-BOX)
\newenvironment{proof}{\noindent {\em Proof.}\quad}{\hfill{\eop}\\}

\date{}

\begin{document} 

\title{Krawtchouk matrices from classical and quantum random walks}

\author{
Philip Feinsilver and Jerzy Kocik\\[.1in]
\small
Department of Mathematics\\
\small
Southern Illinois University\\
\small
Carbondale, IL 62901\\
\small
pfeinsil@math.siu.edu, jkocik@siu.edu
}

%\subjclass{
%05B20, %hadamard
%60G50, %random walk
%47A80, %tensor, operators 
%81P99, %qm computing
%46L53. %noncom probab
%}

%\keywords{Krawtchouk polynomials, Hadamard matrices, 
%symmetric tensors, random walk, quantum computing, quantum probability}

\maketitle

\thispagestyle{empty}

\begin{abstract}
Krawtchouk's polynomials occur classically as orthogonal polynomials
with respect to the binomial distribution. 
They may be also expressed in the form of matrices,
that emerge as arrays of the values that the polynomials take.
The algebraic properties of these matrices provide a very interesting 
and accessible example in the approach to probability theory
known as {\em quantum probability.}
First it is noted how the Krawtchouk matrices are connected to the
classical symmetric Bernoulli random walk. 
And we show how to derive Krawtchouk matrices in the quantum probability
context via tensor powers of the elementary Ha\-da\-mard matrix. 
Then connections with the classical situation are shown by calculating
expectation values in the quantum case.
\end{abstract}

\newpage

%===================  1   ==============================
\section{Introduction}
\label{sec:intro}

Some very basic algebraic rules can be expressed using matrices.
Take, for example,
$$ 
\begin{array}{lcl}  
        (a+b)^2   &=& a^2+2ab+b^2          \\
       (a+b)(a-b) &=& a^2\phantom{+2ab\,\,\,} - b^2           \\
        (a-b)^2   &=& a^2-2ab+b^2  
\end{array}
\qquad\Rightarrow\qquad
K^{(2)} = \left[\begin{array}{rrr}
                             1 &  1 &  1 \cr
                             2 &  0 & -2 \cr
                             1 & -1 &  1 \cr \end{array}\right]
$$
(the expansion coefficients make up the columns of the matrix).
In general, we make the definition:

%----------------------
\begin{definition}\rm
The $N^{\mathrm{th}}$-order Krawtchouk  matrix $\FFF N{}$ is an
$(N+1)\times(N+1)$ matrix,
the entries of which are determined by the expansion:
\begin{equation}
\label{eq:genkraw}
    (1+v)^{N-j} \; (1-v)^j = \sum_{i=0}^{N} \ v^i K^{(N)}_{ij} \,.
\end{equation}
The left-hand-side
$
              G(v)= (1+v)^{N-j}\; (1-v)^j
$
is thus the {\it generating function} for the 
row entries of the $j^{\mathrm{th}}$ column of $K^{(N)}$.
Expanding gives an explicit expression:
$$
K^{(N)}_{ij}= \sum_{k} (-1)^k {j \choose k}
      {N-j \choose i-k}     \,.
$$ 
\end{definition}
Here are the Krawtchouk  matrices of order zero and one:
\begin{equation}
\label{eq:krav2}
K^{(0)}=\left[\begin{array}{rr}{ 1 }\end{array}\right] \qquad
K^{(1)}=\left[        
  \begin{array}{rr}
             1 &  1 \cr
             1 & -1 \cr 
             \end{array}\right]  \,.
 \end{equation}
More examples can be found in Table 1 of Appendix 1. 
In the remaining of the text, matrix indices run from $0$ to $N$.
\\

One may view the columns of Krawtchouk matrices 
as  \emph{generalized binomial coefficients}.
The rows define Krawtchouk \emph{polynomials}: for a fixed order 
$N$, the $i$-th \emph{Krawtchouk polynomial} is the function 
$$
        K_i(j,N) = K^{(N)}_{ij}   
$$ 
that takes its corresponding values from the $i$-th row. 
One can easily show that $K_i(j,N)$ is indeed a 
polynomial of degree $i$ in the variable $j$. 
\\

Historically, Krawtchouk's polynomials were introduced and studied 
by Mikhail Krawtchouk in the late 20's \cite{Kra, Kra2}. %1929,33
Since then, they have appeared in many areas of mathematics and applications.
As orthogonal polynomials, they occur in the classic work by
Sz\"ego \cite{Sze}. %1959
They have been studied from the point of view of 
harmonic analysis and special functions, 
e.g., in work of Dunkl \cite{Dun, DR}. %1974,76
In statistical considerations, they arose in work of 
Eagleson \cite{Eag} %1969 
and later Vere-Jones \cite{V-J}. %1971
They play various roles in coding theory and combinatorics, for example, in
MacWilliams' theorem on weight enumerators \cite{MS, Lev}, %1975, 97
and in association schemes \cite{Del1,Del2,Del3}.   %1972,73
\\

A classical probabilistic interpretation has been given in \cite{FS}. %1991
In the context of the classical symmetric random walk,
it is recognized that Krawtchouk's polynomials 
are elementary symmetric functions in variables taking values
$\pm1$. Specifically, if $\xi_i$ are independent Bernoulli random
variables taking values $\pm1$ with probability $\frac12$, then if
$j$ of the $\xi_i$ are equal to $-1$, the $i^{\rm th}$ elementary
symmetric function in the $\xi_i$ is equal to $K^{(N)}_{ij}$.
It turns out that the generating function (\ref{eq:genkraw}) is a
martingale in the parameter $N$. Details are in Section
\ref{sec:classical} below.
\\

As matrices, they appeared in the 1985 work of N. Bose \cite{B} 
on digital filtering, in the context of the Cayley
transform on the complex plane. 
The symmetric version of the Krawtchouk matrices has been considered
in \cite{FF}. 
\\

Despite this wide research, the full potential, meaning and significance 
of Krawtchouk polynomials is far from being complete. 
In this paper we look at Krawtchouk matrices as \emph{operators} and 
propose two new ways in which Krawtchouk matrices arise:
via classical and quantum random walks.
Especially the latter is of current interest.
The starting idea is to represent the second Krawtchouk matrix
(coinciding with the basic Hadamard matrix) as a sum of two operators
$$ 
\left[\begin{array}{rr}
           1 &  1 \cr
           1 & -1 \cr 
\end{array}\right]
=
\left[\begin{array}{cc}
           0 & 1\cr 
           1 & 0\cr
\end{array}\right] 
+
\left[\begin{array}{rr}
          1 & 0\cr 
          0 &-1\cr
\end{array}\right] .
$$
Via the technique of tensor products of underlying spaces we obtain
a relationship between Krawtchouk matrices and Sylvester-Hadamard
matrices.
\\

The reader should consult Parthasarathy's \cite{Par} for 
material on quantum probability. It contains the operator theory
needed for the subject as well as showing the connections with 
classical probability theory.\\

For information on Hadamard matrices as they appear here, we recommend 
Yarlagadda and Hershey's work  \cite{YH} which provides 
an overview of the subject of Sylvester-Hadamard matrices, 
indicating many interesting applications. %1997
For statisticians, they point out that in Yates' factorial analysis, 
the Hadamard transform provides a useful nonparametric test for association.
\\

Yet another area of significance of this research
lies in the quantum computing program \cite{Lom, Par}.
Details on this connection will appear in an independent work.
\\

This paper is organized as follows.
In the next section, we review basic properties 
of Kraw\-tchouk matrices. The identities presented, although
basic, seem to be new and do not appear in the references cited. 
Section \ref{sec:classical} presents the classical
probability interpretation.  It may be viewed as a warm-up 
leading to the \emph{quantum random walk} 
introduced and studied in Section \ref{sec:quantum}, 
and to the relationship between Krawtchouk matrices 
and Sylvester-Hadamard matrices. 
The generating function techniques used there are original with the
present authors. In the last subsection,
calculating expectation values in the quantum case shows how the
quantum model is related to the classical random walk.
Appendix~1 has examples of Krawtchouk and
symmetric Krawtchouk matrices so that the reader may see concretely the
subject(s) of our discussion.
Appendices~2 (tensor products) and 3 (symmetric tensor spaces) 
are included to aid the reader as well as to clarify the notation.
\\  

%===================  2   ===========================================
\section{Basic properties of Krawtchouk matrices}
\label{sec:ident}

\noindent
(1)
The square of a Krawtchouk  matrix is proportional to the identity matrix.
$$
(\FFF N{})^2 = 2^N\,I   \,.
$$
This remarkable property allows one to define 
a Fourier-like {\em Krawtchouk transform} on integer vectors. 
\\
\\
(2)
The top row is all 1's. The bottom row has $\pm 1$'s with
alternating signs, starting with $+1$.
The leftmost entries are just binomial coefficients,
$\FFF N{i0} = {N\choose i}\phantom{\biggm|}$.  
The rightmost entries are binomial
coefficients with alternating signs, 
$\FFF N{iN} = (-1)^i{N\choose i}\phantom{\biggm|}$.
\\
\\
(3)
There is a four-fold symmetry:
$
|\FFF{N}{i\;  j}|  =  |\FFF{N}{N-i\;j}| =
|\FFF{N}{i\;N-j} | =  |\FFF{N}{N-i\;N-j}|
$.\\

Krawtchouk matrices generalize Pascal's triangle in the following sense:
Visualize a stack of Krawtchouk matrices, the order $N$ increasing
downwards. 
Pascal's triangle is formed by the leftmost columns.
It turns out that Pascal's identity
holds for the other columns as well.
Less obvious is another identity --- call it dual Pascal.

\begin{proposition}
\label{prop:ids}
Set
$ a= \FFF N {i-1\;j},\,
  b= \FFF N {i \;j}$,
$ A= \FFF {N+1} {i\;j},\,
  B= \FFF {N+1} {i\;j+1}$ . \\

1. {\bf (Cross identities)}
The following mutually inverse relations (Pascal and dual Pascal) hold:
\begin{displaymath}
\begin{array}{cc}
a+b=A  \\
b-a=B  \\
\end{array} 
\qquad\hbox{and}\qquad
\begin{array}{cc}
A+B=2b \\
A-B=2a \,.
\end{array} 
\end{displaymath}

2. {\bf (Square identity)}
In a square of any four adjacent entries in a Krawtchouk  matrix, 
the entry in the left-bottom corner is the sum of the other three,
i.e., 
$$
\hbox{for } ~~ 
K =  
\left[\begin{array}{cccc}
              ~  &\vdots&\vdots &\cr
              \cdots & a & c &\cdots \cr
              \cdots & b & d &\cdots\cr
               ~ &\vdots&\vdots &\cr
\end{array}\right]
\qquad\hbox{one has}\qquad
b=a+c+d.
$$

\end{proposition}

\begin{proof}
For $a+b$, consider 
$$(1+v)^{N+1-j}(1-v)^j=(1+v)\, (1+v)^{N-j}(1-v)^j  \,.$$
For $b-a$, consider 
$$(1+v)^{N-j}(1-v)^{j+1}= (1-v) \,  (1+v)^{N-j}(1-v)^j \,. $$
The inverse relations are immediate. 
The square identity follows from the observation
$(a+b)+(c+d) = A+B=2b$, hence $a+c+d=b$.
\end{proof}

The square identity is useful in producing the entries of a Krawtchouk 
matrix: fill the top row with 1's, the right-most column with 
sign-alternating binomial coefficients. Then, apply the square identity to 
reproduce the matrix. 
\\

In summary, the identities considered above can be written as follows:
\\

{\bf Cross identities:}
$$
\begin{array}{ccccc}
(i)&   \FFF N{i-1\;j} + \FFF N{i  \;j}  =  \FFF {N+1}{i\;j}    &\quad&
(ii)&  \FFF N{i  \;j} + \FFF N{i  \;j+1}= 2\FFF {N-1}{i\;j}   \cr\mathstrut\cr
(iii)& \FFF N{i \;j}  - \FFF N{i-1\;j}  =  \FFF {N+1}{i\;j+1}  &\quad&
(iv)&  \FFF N{i  \;j} - \FFF N{i  \;j+1}= 2\FFF {N-1}{i-1\;j}\,.\cr
\end{array}
$$
\\

{\bf Square identity:}
$$
\FFF N {ij} =  \FFF N{i-1\;j} + \FFF N{i-1\; j+1}+  \FFF N{i\;j+1} \,.
$$
\\

If each column of the matrix is multiplied by the corresponding binomial
coefficient, the matrix becomes symmetric.
Let $B^{(N)}$ denote the $(N+1)\times (N+1)$ diagonal matrix 
with binomial coefficients 
\begin{equation}
\label{eq:diagB}
          B^{(N)}_{ii}={N\choose i}
\end{equation}
as its non-zero entries.
Then, for each $N\ge0$, 
one defines the  {\bf symmetric Krawtchouk  matrix} as
$$
S^{(N)}=\FFF N{}B^{(N)} \,.
$$
{\bf Example:}  For $N=3$, we have
$$
S^{(3)} =
       \left[\begin{array}{rrrr} 
                      1 &  1 &  1  &  1 \cr
                      3 &  1 & -1  & -3 \cr
                      3 & -1 & -1  &  3 \cr
                      1 & -1 &  1  & -1 \cr \end{array}\right]
       \left[\begin{array}{rrrr} 
                      1 &  0 & 0 & 0 \cr
                      0 &  3 & 0 & 0 \cr
                      0 &  0 & 3 & 0 \cr
                      0 &  0 & 0 & 1 \cr \end{array}\right]
=
      \left[\begin{array}{rrrr} 
                     1 &  3 &  3  &  1 \cr
                     3 &  3 & -3  & -3 \cr
                     3 & -3 & -3  &  3 \cr
                     1 & -3 &  3  & -1 \cr \end{array}\right] \,.
$$ 
Some symmetric Krawtchouk  matrices are displayed in Table 2 of
Appendix 1.  
\\

%============================ 3 ==========================
\section{Krawtchouk matrices and classical random walk}
\label{sec:classical}

In this section we will give a probabilistic meaning to
the Krawtchouk matrices and some of their properties.
\\

Let $\xi_i$ be independent symmetric Bernoulli random variables taking
values $\pm1$. Let $X_N=\xi_1+\cdots+\xi_N$ be the associated random
walk starting from $0$. Now observe that the generating function of
the elementary symmetric functions in the $\xi_i$ is a martingale, in
fact a discrete exponential martingale:
$$
   M_N = \prod_{i=1}^N(1+v\xi_i)=\sum_k v^k 
           \alpha_k(\xi_1,\ldots,\xi_N)   \,,
$$
where $\alpha_k$ denotes the $k^{\mathrm{th}}$ elementary symmetric
function. 
The martingale property is immediate since each $\xi_i$ has mean $0$.
Suppose that at time $N$, the number of the $\xi_i$ that are equal to $-1$
is $j_N$, with the rest equal to $+1$. Then  $j_N= (N-X_N)/2$ 
and $M_N$ can be expressed solely in terms of $N$ and $X_N$, or, 
equivalently, of $N$ and $j_N$
$$
       M_N = (1+v)^{N-j_N}(1-v)^{j_N} 
           = (1+v)^{(N+X_N)/2}(1-v)^{(N-X_N)/2}   \,.
$$
 From the generating function for the Krawtchouk
matrices, (\ref{eq:genkraw}), follows 
$$
    M_N = \sum_iv^iK^{(N)}_{i,j_N} \,,
$$
so that as functions on the Bernoulli space, each sequence of random
variables $K^{(N)}_{i,j_N}$ is a martingale. 
\\

Now we can interpret two basic recurrences of Proposition \ref{prop:ids}.
For a fixed column of $K^{(N)}$, the corresponding column in 
$K^{(N+1)}$ satisfies the Pascal triangle recurrence:
$$
  \FFF N{i-1\;j} + \FFF N{i  \;j}  =  \FFF {N+1}{i\;j} \,.
$$
To see this in the probabilistic setting, write
$M_{N+1}=(1+v\xi_N)M_N$.  Observe that for $j_N$ to remain constant,
$\xi_N$ must take the value $+1$ and expanding
$(1+v)M_N$ yields the Pascal recurrence as in the proof of Proposition
\ref{prop:ids}.
It is interesting how the martingale property comes into play.
We have
$$
\FFF N{ij_N}= E(\FFF {N+1}{ij_{N+1}}|\xi_1,\ldots,\xi_N)
             =\frac12\,\left(\FFF{N+1}{i\,j_N+1}+\FFF{N+1}{ij_N}\right)\,,
$$
since half the time $\xi_{N+1}$ is $-1$, increasing $j_N$ by 1, and
half the time $j_N$ is unchanged. Thus, writing $j$ for $j_N$,
$$
 \FFF N{ij}=\frac12\,\left(\FFF{N+1}{i\,j+1}+\FFF{N+1}{ij}\right) \,.
$$

Many further properties of Krawtchouk polynomials 
may be derived from their interpretation as elementary symmetric
functions on the Bernoulli space with scope for probabilistic methods
as well.
\\

%\vfill
%\newpage  %<<<<<<<

%========================= 4 ======================
\section{Krawtchouk matrices and quantum random walk}
\label{sec:quantum}

In quantum probability, random variables are modeled by self-adjoint
operators on Hilbert spaces and independence by tensor products. 
We can model a symmetric Bernoulli random walk as follows. 
Consider a 2-dimensional Hilbert space $V={\bf R}^2$
and two special $2\times2$ operators,
$$ 
    F=\left[\begin{array}{cc}0&1\cr 1&0\cr\end{array}\right]\qquad\hbox{and}\qquad
    G=\left[\begin{array}{rr}1&0\cr  0&-1\cr\end{array}\right]\,,
$$
satisfying $F^2=G^2=I$ (the $2\times2$ identity).
The fundamental Ha\-da\-mard matrix $H$ coincides with the second
Kraw\-tchouk matrix.  Now we shall view it as a sum of the above operators 
%equations (\ref{eq:krav2}) and (\ref{eq:had})
$$
H=F+G = \left[\begin{array}{rr} 1 &  1 \cr
                       1 & -1 \cr\end{array}\right] \,.
$$
One can readily check that
\begin{equation}
\label{eq:H}
         FH =F(F+G) =(F+G)G =HG
\end{equation}
(use $F^2=G^2=I$). This, of course, can be viewed as 
the spectral decomposition of $F$ and we can interpret the Hadamard
matrix as a matrix reducing $F$ to diagonal form. 

\begin{remark}\rm \label{rem:exval}
Note that the  exponentiated  operator
$$
    \exp(zF) = \left[\begin{array}{cc}\cosh z&\sinh z\cr \sinh z&\cosh
z\cr\end{array}\right]
$$
has the expectation value in the state $e_0$ equal to
\begin{equation}\label{eq:exval}
\langle e_0,\exp(zF)e_0\rangle = \cosh z\,,
\end{equation}
where $e_0$ denotes the transpose of $[1,0]$.
This coincides with the moment generating function for the symmetric 
Ber\-noul\-li random variable taking values $\pm 1$,
showing that indeed we are dealing with the (quantum)
generalization of the classical model.
\end{remark}

The Hilbert space of states is represented 
by the $N$-th tensor product of the original space $V$,
that is, by the $2^N$-dimensional Hilbert space $V^{\otimes N}$ \ 
(see Appendix~2 for notation). 
Define the following linear operators, $N$ in all, in $V^{\otimes N}$
\begin{eqnarray*}
f_1 &=& F\otimes I \otimes\cdots\otimes I          \\
f_2 &=& I\otimes F\otimes I \otimes\cdots\otimes I \\
\vdots &=& \vdots                                    \\
f_N &=& I\otimes I \otimes\cdots\otimes F  \,,
\end{eqnarray*}
each $f_i$ describing a flip at the $i$-th position.
These are the quantum equivalents of the random walk variables 
from Section \ref{sec:classical}.
We shall consider the superposition of these independent 
actions,  setting 
$$
  X_F=f_1+\cdots+f_N \,.
$$
%
%{\bf begin{notation}\rm
{\bf Notation:}
For notational clarity,
since $N$ is fixed throughout the discussion, we drop the index $N$ from 
the $X$'s. 
%\end{notation}
\\

Analogously, we define:
\begin{eqnarray*}
g_1 &=& G\otimes I \otimes\cdots\otimes I\\
g_2 &=& I\otimes G\otimes I \otimes\cdots\otimes I\\
\vdots &=& \vdots \\
g_N &=& I\otimes I \otimes\cdots\otimes G \,,
\end{eqnarray*}
with $X_G=g_1+\cdots+g_N$. 
Finally, let us extend $H$ to the $N$-fold tensor product, 
setting $H_N=H^{\otimes N}$. These are the well-known Sylvester-Hadamard
matrices with the first few given here:
\def\b{\circ}   \def\a{\bullet}
$$
H_{1}=\left[\begin{array}{cc}
\a &\a  \cr
\a &\b  \cr
\end{array}\right]
\quad
H_{2}=\left[\begin{array}{cccc}
\a &\a &\a &\a  \cr
\a &\b &\a &\b   \cr 
\a &\a &\b &\b   \cr
\a &\b &\b &\a   \cr
\end{array}\right]
\quad
H_{3}=\left[\begin{array}{cccccccc}
\a &\a &\a &\a  &\a &\a &\a &\a    \cr
\a &\b &\a &\b  &\a &\b &\a &\b    \cr
\a &\a &\b &\b  &\a &\a &\b &\b    \cr
\a &\b &\b &\a  &\a &\b &\b &\a    \cr
\a &\a &\a &\a  &\b &\b &\b &\b    \cr 
\a &\b &\a &\b  &\b &\a &\b &\a    \cr 
\a &\a &\b &\b  &\b &\b &\a &\a    \cr 
\a &\b &\b &\a  &\b &\a &\a &\b    \cr 
\end{array}\right] \,,
$$ 
etc.,  
where, for typographical reasons, we use
$\bullet$ for $1$ and $\circ$ for $-1$.
\\

It turns out that our $X$-operators intertwine
the Sylvester-Hadamard matrices.  
For illustration, consider a calculation for $N=3$:
\begin{eqnarray*}
f_1H_3 &=& (F\otimes I\otimes I)(H\otimes H\otimes H)\\
         &=& (H\otimes H\otimes H)(G\otimes I\otimes I)
              = H_3g_1 \,,
\end{eqnarray*}
where the relation $FH=HG$ is used. 
This clearly generalizes to $f_kH_N=H_Ng_k$ and, 
by summing over $k$, yields an important relation:
$$
            X_F H_N = H_N X_G \,.
$$
Now, we shall consider the symmetrized versions of the operators
(the reader is referred to Appendix 3 for the theory and methods used here).
Since products are preserved in the process of reduction to
the symmetric tensor space, we get
$$
       \overline{X}_F\overline{H}_N = \overline{H}_N \overline{X}_G \,,
$$
the bars indicating the corresponding induced maps.
We know how to calculate $\overline{H}_N$ from the action of $H$ on
polynomials in degree $N$. For symmetric tensors the components in 
degree $N$ are
$$ 
      x_0^{N-k}x_1^k \,,
$$
where $0\le k\le N$.

\begin{proposition} 
For each $N>0$, symmetric reduction of the $N^{\rm th}$ Hadamard 
matrix results in the transposed $N^{\rm th}$ Krawtchouk matrix:
$$ 
    (\overline{H}_N)_{ij} = \FFF{N}{ji}  \,.
$$
\end{proposition}

\begin{proof}
Writing $(x,y)$ for $(x_0,x_1)$, we have in degree $N$ for the 
$k^{\mathrm{th}}$ component:
$$
   (x+y)^{N-k}(x-y)^k=\sum_l \overline{H}_{kl} \,\dir l.  \,.
$$
Scaling out $x^N$ and replacing $v=y/x$ yields
the generating function for the Krawtchouk matrices with the
coefficient of $v^l$ equal to $\FFF{N}{lk}$. Thus the result.
\end{proof}

Now consider the generating function for the elementary symmetric
functions in the quantum variables $f_j$. This is the $N$-fold
tensor power
$$ 
     {\mathcal F}_N (t) = (I+tF)^{\otimes N}
                    = I^{\otimes N}+t\,X_F+\cdots  \,,
$$
noting that the coefficient of $t$ is $X_F$. 
Similarly, define
$$
     {\mathcal G}_N(t) = (I+tG)^{\otimes N}
                   = I^{\otimes N}+tX_G+\cdots \,.
$$
  From $(I+tF)H=H(I+tG)$ we have
$$
   {\mathcal F}_NH_N  =  H_N{\mathcal G}_N        
   \qquad\hbox{and}\qquad 
   \overline{\mathcal F}_N \overline{H}_N  
 = \overline{H}_N \overline{\mathcal G}_N  \,.
$$
The difficulty is to calculate the action on the symmetric tensors for
operators, such as $X_F$, that are not pure tensor powers. However,
from ${\mathcal F}_N(t)$ and ${\mathcal G}_N(t)$ we can 
recover $X_F$ and $X_G$ via
$$
          X_F=\frac{d}{dt}\biggm|_{t=0}(I+tF)^{\otimes N},
     \qquad
          X_G=\frac{d}{dt}\biggm|_{t=0}(I+tG)^{\otimes N}
$$
with corresponding relations for the barred operators.
Calculating on polynomials yields the desired results as follows.
$$ 
          I+tF = \left[\begin{array}{cc}1&t\cr t&1\cr\end{array}\right],\qquad
          I+tG = \left[\begin{array}{cc}1+t&0\cr 0&1-t\cr\end{array}\right] \,.
$$
In degree $N$, using $x$ and $y$ as variables, we get the 
$k^{\mathrm{th}}$ component for $\overline{X}_F$ and $\overline{X}_G$
via
\begin{eqnarray*}
    \frac{d}{dt}\biggm|_{t=0}(x+ty)^{N-k}(tx+y)^k
                     &=& (N-k)\,\dirs k+1. +k\,\dirs k-1. \,, 
                   \end{eqnarray*}
and since $I+tG$ is diagonal,
\begin{eqnarray*}
    \frac{d}{dt}\biggm|_{t=0}(1+t)^{N-k}(1-t)^k\,\dir k.
                     &=& (N-2k)\,\dir k.   \,. 
\end{eqnarray*}
For example, calculations for $N=4$ result in
\begin{eqnarray*}
\overline{X}_F = \left[
         \begin{array}{ccccc}0&4&0&0&0\cr 
                  1&0&3&0&0\cr 
                  0&2&0&2&0\cr 
                  0&0&3&0&1\cr
                  0&0&0&4&0\cr\end{array}\right],&\qquad&
\overline{H}_4 = \left[\begin{array}{rrrrr}
                    1&4&6&4&1\cr
                    1&2&0&-2&-1\cr 
                    1&0&-2&0&1\cr
                    1&-2&0&2&-1\cr
                    1&-4&6&-4&1\cr \end{array} \right]\,,\\ 
\mathstrut\\
\overline{X}_G  = \left[\begin{array}{rrrrr}4&0&0&0&0\cr
                     0&2&0&0&0\cr
                     0&0&0&0&0\cr
                     0&0&0&-2&0\cr
                     0&0&0&0&-4\cr\end{array}\right] \,.&&
\end{eqnarray*}
Since $\overline{X}_G$ is the result of diagonalizing
$\overline{X}_F$, we observe that

\begin{corollary}
The spectrum of $\overline{X}_F$ is
$N,N-2,\ldots,2-N,-N$, coinciding with the support 
of the classical random walk.
\end{corollary}

\bigskip

%----------------------------------------------------
\subsection{Expectation values}
%For quantum variables, spectral measures provide classical
%probability distributions and hence classical interpretations. 
To find the probability distributions associated to our $X_F$ operators, 
we must calculate expectation values, cf. Remark \ref{rem:exval}. In the present context, 
expectation values in two particular states are especially interesting. 
Namely, in the state $e_0$ and in the normalized trace, which is the uniform 
distribution on the spectrum. In the $N$-fold tensor product, 
we want to consider expectation values in the ground state
$|\,000\ldots0\,\rangle$ and normalized traces. 
Then we can go to the symmetric tensors.
\\

The scalar product on the tensor product space factors, 
corresponding to independence in classical probability. Thus,
from (\ref{eq:exval}) one obtains the expectation value of $\exp(zX_F)$ 
in the ground state $|\,000\ldots0\,\rangle$ to be $(\cosh z)^N$. 
Similarly, the trace of the tensor product of operators is the product
of their traces. So, for the trace, $\tr \exp(zF)=2\cosh z$ implies
$\tr\exp(zX_F)=2^N(\cosh z)^N$ and, after normalizing, this yields
$(\cosh z)^N$.
\\

For the barred operators, we consider the symmetric trace. Here we use
the {\bf symmetric trace theorem}, detailed in Appendix 3. It tells us
that the generating function for the symmetric traces of any operator
$A$ in the various degrees is \\ $\det(I-tA)^{-1}$. Taking
$A=\exp(zF)$, we have
\begin{eqnarray*}
  \det(I-te^{zF})^{-1} &=& [(1-te^z)(1-te^{-z})]^{-1}\\
                       &=& (1-2t\cosh z+t^2)^{-1} \,.
\end{eqnarray*}
The latter is the generating function for Chebyshev polynomials of the
second kind, $U_N$, so that the normalized symmetric trace is
$$
   (N+1)^{-1}\tr_\mathrm{Sym}^N \exp(zF) = U_N(\cosh z)/(N+1) \,,
$$
which equals as well
$$
      \frac{e^{z(N+1)} - e^{-z(N+1)}}{(e^z-e^{-z})(N+1)}
= \frac{\sinh (N+1)z}{(N+1)\,\sinh z} \,.
$$
This corresponds to a uniform distribution on the support of the
random walk at time $N$, namely, $-N,2-N,\ldots,N-2,N$.

\vskip.3in
{\bf Acknowledgment.} We would like to thank Marlos Viana for inviting us
to participate in the special session and we extend our appreciation
for all the hard work involved in organizing the session as well as
related activities.
\\

\newpage

%==================  krav matrices ==================================
\section*{Appendix 1: Krawtchouk matrices}
\bigskip
\hrule
\bigskip

\begin{eqnarray*}
K^{(0)}&=&\left[\begin{array}{r} 1 \end{array}\right]   
\\
\\
K^{(1)}&=&\left[\begin{array}{rr} 1 &  1 \cr
                           1 & -1 \cr\end{array}\right]    
\\
\\
K^{(2)}&=& \left[\begin{array}{rrr}  1 &  1 &  1 \cr
                             2 &  0 & -2 \cr
                             1 & -1 &  1 \cr \end{array}\right]
\\
\\
K^{(3)}&=&
       \left[\begin{array}{rrrr}
                      1 &  1 &  1  &  1 \cr
                      3 &  1 & -1  & -3 \cr
                      3 & -1 & -1  &  3 \cr
                      1 & -1 &  1  & -1 \cr 
             \end{array}\right]
\\
\\
K^{(4)}&=&
       \left[\begin{array}{rrrrr} 1 &  1 &  1  &  1  &  1 \cr
                      4 &  2 &  0  & -2  & -4 \cr
                      6 &  0 & -2  &  0  &  6 \cr
                      4 & -2 &  0  &  2  & -4 \cr
                      1 & -1 &  1  & -1  &  1 \cr \end{array}\right]
\\
\\
K^{(5)}&=&
       \left[\begin{array}{rrrrrr} 1  &  1 &  1  &  1  &  1 &   1 \cr
                      5  &  3 &  1  & -1  & -3 &  -5 \cr
                     10  &  2 & -2  & -2  &  2 &  10 \cr
                     10  & -2 & -2  &  2  &  2 & -10 \cr
                      5  & -3 &  1  &  1  & -3 &   5 \cr
                      1  & -1 &  1  & -1  &  1 &  -1 \cr \end{array}\right]
\\
\\
K^{(6)}&=&
       \left[\begin{array}{rrrrrrr} 
                      1  &  1 &  1  &  1  &  1 &  1  &   1 \cr
                      6  &  4 &  2  &  0  & -2 & -4  &  -6 \cr
                     15  &  5 & -1  & -3  & -1 &  5  &  15 \cr
                     20  &  0 & -4  &  0  &  4 &  0  & -20 \cr
                     15  & -5 & -1  &  3  & -1 & -5  &  15 \cr
                      6  & -4 &  2  &  0  & -2 &  4  &  -6 \cr
                      1  & -1 &  1  & -1  &  1 & -1  &   1 \cr
\end{array}\right]
\end{eqnarray*}

\bigskip
\hrule
\bigskip
\centerline{{\bf Table 1:} Krawtchouk  matrices}

%=============   symmetric krav matrices =============================
\newpage
~
\hrule
\bigskip

\begin{eqnarray*}
S^{(0)}&=&\left[\begin{array}{r} 1 \end{array}\right]
\\
\\
S^{(1)}&=&\left[\begin{array}{rr} 1 &  1 \cr
                        1 & -1 \cr\end{array}\right]
\\
\\
S^{(2)}&=& \left[\begin{array}{rrr}  1 &  2 &  1 \cr
                          2 &  0 & -2 \cr
                          1 & -2 &  1 \cr \end{array}\right]
\\
\\
S^{(3)}&=&
      \left[\begin{array}{rrrr} 1 &  3 &  3  &  1 \cr
                     3 &  3 & -3  & -3 \cr
                     3 & -3 & -3  &  3 \cr
                     1 & -3 &  3  & -1 \cr \end{array}\right]
\\
\\
S^{(4)}&=&
      \left[\begin{array}{rrrrr} 1 &  4 &  6  &  4  &  1 \cr
                     4 &  8 &  0  & -8  & -4 \cr
                     6 &  0 &-12  &  0  &  6 \cr
                     4 & -8 &  0  &  8  & -4 \cr
                     1 & -4 &  6  & -4  &  1 \cr \end{array}\right]
\\
\\
S^{(5)}&=&
      \left[\begin{array}{rrrrrr} 1  &  5 & 10 &  10  &   5 &   1 \cr
                     5  & 15 & 10 & -10  & -15 &  -5 \cr
                    10  & 10 &-20 & -20  &  10 &  10 \cr
                    10  &-10 &-20 &  20  &  10 & -10 \cr
                     5  &-15 & 10 &  10  & -15 &   5 \cr
                     1  & -5 & 10 & -10  &   5 &  -1 \cr \end{array}\right]
\\
\\
S^{(6)}&=&
       \left[\begin{array}{rrrrrrr} 1  &   6 &  15  &  20 &  15 &   6 &   1
\cr
                      6  &  24 &  30  &   0 & -30 & -24 &  -6 \cr
                     15  &  30 & -15  & -60 & -15 &  30 &  15 \cr
                     20  &   0 & -60  &   0 &  60 &   0 & -20 \cr
                     15  & -30 & -15  &  60 & -15 & -30 &  15 \cr
                      6  & -24 &  30  &   0 & -30 &  24 &  -6 \cr
                      1  &  -6 &  15  & -20 &  15 &  -6 &   1 \cr
\end{array}\right]
\end{eqnarray*}

\bigskip
\hrule
\bigskip
\centerline{{\bf Table 2:} Symmetric Krawtchouk  matrices}

%====================== A1 ==========================
\section*{Appendix 2: Tensor products}

Fulton and Harris \cite{FH} is a useful reference for
this section and the next. Also Parthasarathy \cite{Par}, Chapter II, is an
excellent reference.
\\

Let $V$ be a $d$-dimensional vector space over ${\bf R}$. 
We fix an orthonormal basis $\{e_0,\ldots,e_\delta\}$
with $d=1+\delta$. Denote tensor powers of $V$ by
$V^{\otimes N}$, so that $V^{\otimes 2}=V\otimes V$, etc. 
A basis for $V^{\otimes N}$ is given by all $N$-fold tensor products 
of the basis vectors $e_i\,$,
$$ 
|\,n_1n_2\ldots n_N\,\rangle = 
      e_{n_1}\otimes e_{n_2}\otimes \cdots\otimes e_{n_N} \,.
$$
Note that we can label these $d^N$ basis elements by all  
numbers $0$ to $d^N-1$ and recover the tensor products 
by expressing these numbers in base $d$, 
putting leading zeros so that all extended labels are of length $N$. 
\\

Now let $\{A_i:\; 1\le i\le N\}$ be a set of $N$ linear operators on $V$. 
On $V^{\otimes N}$, the linear operator 
$A=A_1\otimes A_2\otimes\cdots\otimes A_N$ acts on a basis vector
$|\,n_1n_2\ldots n_N\,\rangle$ by
$$ 
 A |\,n_1n_2\ldots n_N\,\rangle =
       A_1e_{n_1}\otimes\cdots\otimes A_Ne_{n_N} \,.
$$
This needs to be expanded and terms regrouped using bilinearity.
\\

If $A$ and $B$ are two $d\times d$ matrices, the matrix corresponding
to the operator $A\otimes B$ is the Kronecker product, a
$d^2\times d^2$ matrix having the block form:
$$\left[\begin{array}{ccc}
       a_{00}B&\ldots&a_{0\delta}B\cr
       a_{10}B&\ldots&a_{1\delta}B\cr
              \vdots&\vdots&\vdots\cr
       a_{\delta0}B&\ldots&a_{\delta\delta}B\cr\end{array}\right] \,.
$$
This, iteratively, is valid  for higher-order tensor
products (associating from the left by convention).
The rows and columns of the matrix of a linear operator acting on
$V^{\otimes N}$ are conveniently labeled by associating to each 
basic tensor
$|\,n_1n_2\ldots n_N\,\rangle$ the corresponding integer label
$\sum\limits_{k=1}^N n_kd^{N-k}$, which thus provides a canonical
ordering.
\\

%======================= A2 ============================
\section*{Appendix 3: Symmetric tensor spaces}

Here we review symmetric tensor spaces
as spaces of polynomials in commuting variables.
This material is presented with a view to the infinite-dimensional
case in \cite{Par}, pp. 105ff., 
however we focus on the finite-dimensional context and include as well
an important observation contained in Theorem \ref{thm:symmtr}.
\\

The space $V^{\otimes N}$ can be mapped onto the space of symmetric
tensors, $V^{\otimes_S N}$ by identifying basis vectors 
(in $V^{\otimes N}$) that are equivalent under all permutations. 
Alternatively, one can identify the basic tensor
$|\,n_1n_2\ldots n_N\,\rangle$ with the monomial $x_{n_1}x_{n_2}\cdots
x_{n_N}$
in the commuting variables $x_0,\ldots,x_\delta$. 
Hence we have a linear map
from tensor space into the space of polynomials, itself 
isomorphic to the space of symmetric tensors:
$$
  \overline{\phantom{W}}:\quad \bigcup_{N\ge0}V^{\otimes N} 
                \ \longrightarrow \ 
                {\bf R}[x_0,\ldots,x_\delta]  
              \cong \bigcup_{N\ge0} V^{\otimes_S N} \,.
$$
In the symmetric tensor space, tensor labels
need to count only occupancy, that is, the number of times a
basis vector of $V$ occurs in a given basic tensor of $V^{\otimes N}$. 
We indicate occupancy by a multi-index which is the exponent of the
corresponding monomial. The dimension of $V^{\otimes_S N}$ is thus
$$
     \hbox{dim}\; V^{\otimes_S N} = {N+d-1\choose d-1}  \,,
$$
that is, the number of monomials homogeneous of degree $N$. 
\\

Given an operator $A$ on $V$, let $A_N=A^{\otimes N}$. Then $A_N$  
induces an operator $\overline{A}_N$ on $V^{\otimes_S N}$ from the action of
$A$ on polynomials,  which we call the {\bf symmetric representation 
of $A$ in degree $N$}.  For convenience we work dually with the tensor
components rather with the action on the basis vectors.
Denote the matrix elements of the action of
$\overline{A}_N$ by $\overline{A}_{mn}$. If $A$ has matrix entries
$A_{ij}$, let
$$
       y_i=\sum_j A_{ij}x_j \,.
$$
Then the matrix elements of the symmetric representation
are defined by the relation (expansion):
$$
   y_0^{m_0}\cdots y_\delta^{m_\delta}=
               \sum_n \overline{A}_{mn}
                     x_0^{n_0}\cdots x_\delta^{n_\delta}
$$
with multi-indices $m$ and $n$.
\\

Composition of $A_1$ with $A_2$ shows that mapping 
to the symmetric representation is an algebra homomorphism, i.e., 
$$
\overline{A_1A_2}=\overline{A}_1\overline{A}_2  \,.
$$
Explicitly, in basis notation
$$ 
\overline{(A_1A_2)}_{mn}=
          \sum_r (\overline{A_1})_{mr}
                 (\overline{A_2})_{rn}  \,.
$$ 
Define the {\bf symmetric trace} in degree $N$ of $A$ as the trace of
the matrix elements of $\overline{A}_N$, i.e., the sum of the diagonal
matrix elements:
$$
     \tr_{\rm Sym}^N A=\sum_{|m|=N}\overline{A}_{mm}
$$
with $|m|$ denoting, as usual, the sum of the components of $m$.
Observe that if $A$ is upper-triangular, 
with eigenvalues $\lambda_1,\ldots,\lambda_d$, then the trace of
this action on the space of polynomials homogeneous of degree $N$ is 
exactly  $h_N(\lambda_1,\ldots,\lambda_d)$, the $N^{\mathrm{th}}$ 
homogeneous symmetric function in the $\lambda$'s. 
\\

We recall a useful theorem on calculating the symmetric trace.
Since the mapping from $A$ to $\overline{A}_N$ is a homomorphism, a
similarity transformation on $A$ extends to one on $\overline{A}_N$ thus
preserving traces. Now, any matrix is similar to an upper-triangular 
one with the same eigenvalues, thus follows \cite{Spr}:

\begin{thm}{\bf Symmetric trace theorem}
\label{thm:symmtr}
Denoting by $\mathrm{tr}_{\mathrm{Sym}}^N$ the trace of the symmetric
representation on polynomials homogeneous of degree $N$, 
$$
     \frac{1}{\det(I-tA)}=\sum_{N=0}^\infty t^N\tr_{\mathrm{Sym}}^N A \,.
$$
\end{thm}

\begin{proof} 
With $\{\lambda_i\}$ denoting the eigenvalues of $A$,
\begin{eqnarray*}
\frac{1}{\det(I-tA)} 
           &=&\prod_i\frac{1}{1-t\lambda_i}
            = \sum_{N=0}^\infty t^Nh_N(\lambda_1,\ldots,\lambda_d)\cr
           &=&\sum_{N=0}^\infty t^N\tr_{\mathrm{Sym}}^N A  \,,\cr 
\end{eqnarray*}
as stated above.
\end{proof}

\begin{remark}\rm  Note that this result is equivalent to 
{\sl MacMahon's Master Theorem} in combinatorics \cite{Mac}.
\end{remark}

\begin{remark}\rm
  From another point of view, Chen and Louck \cite{CL}, considering 
powers of the  bilinear form $\sum_{i,j} x_iA_{ij}y_j$ rather than
just the linear form as done here, study representation functions that
are analogs of our symmetric Krawtchouk matrices. 
Their $L_{\alpha,\beta}$ is our symmetric representation 
scaled by multinomial factors.
In addition they suggest further interesting generalizations beyond
symmetric tensors.
\end{remark}

%\newpage

%==================================================================


\begin{thebibliography}{10}


\bibitem{B}
N. Bose,
\emph{Digital filters: theory and applications},
North-Holland, 1985.

\bibitem{CL}
W.Y.C. Chen and J.D. Louck, 
\emph{The combinatorics of a class of representation functions,} 
Adv. in Math., {\bf 140} (1998), 207--236. 

\bibitem{Del1}
P. Delsarte,
\emph{Bounds for restricted codes, by linear programming,}
Philips Res. Reports, {\bf 27} (1972) 272--289.

\bibitem{Del2}
P. Delsarte,
\emph{Four fundamental parameters of a code and their combinatorial
significance,}
{Info. \&\ Control}, {\bf 23} (1973) 407--438.

\bibitem{Del3}
P. Delsarte,  
\emph{An algebraic approach to the association schemes of coding theory,}
{Philips Research Reports Supplements}, No. 10, 1973.

\bibitem{Dun}
C.F. Dunkl,
\emph{A Krawtchouk polynomial addition theorem and  wreath products of
symmetric
groups,} 
{Indiana Univ. Math. J.},  {\bf 25} (1976) 335--358.

\bibitem{DR}
C.F. Dunkl and D.F. Ramirez,
\emph{Krawtchouk polynomials and the symmetrization of hypergraphs,}
{SIAM J. Math. Anal.}, {\bf 5} (1974) 351--366.

\bibitem{Eag}
G.K. Eagelson,
\emph{A characterization theorem for positive definite sequences of the
Krawtchouk polynomials,}
{Australian J. Stat}, {\bf 11} (1969) 29--38.

\bibitem{FF}
P. Feinsilver and R. Fitzgerald,
\emph{The spectrum of symmetric Krawtchouk matrices,}
{Lin. Alg. \& Appl.}, {\bf 235} (1996) 121--139. 

\bibitem{FS}
P. Feinsilver and R. Schott,
\emph{Krawtchouk polynomials and finite probability theory,}
{Probability Measures on Groups X}, Plenum, 1991, pp. 129--135.

\bibitem{FH}
W. Fulton and J. Harris,
\emph{Representation theory, a first course,}
Graduate texts in mathematics, {\bf 129},
Springer-Verlag, 1991.

\bibitem{Kra}
M. Krawtchouk,
\emph{Sur une generalisation des polynomes d'Hermite,}
{Comptes Rendus}, {\bf 189} (1929) 620--622.

\bibitem{Kra2}
M. Krawtchouk, 
\emph{Sur la distribution des racines des polynomes orthogonaux,}
{Comptes Rendus}, {\bf 196} (1933) 739--741.

\bibitem{Lev}
V.I. Levenstein,
\emph{Krawtchouk polynomials and universal bounds for codes and design in
Hamming spaces,}
{IEEE Transactions on Information Theory}, {\bf 41} 5 (1995)  
1303--1321.

\bibitem{Lom}
S.J. Lomonaco, Jr., 
\emph{A Rosetta Stone for quantum mechanics with an introduction
to quantum computation,}
http://www.arXiv.org/abs/quant-ph/0007045

\bibitem{Mac}
P.A. MacMahon,
\emph{Combinatory analysis,}
Chelsea, New York, 1960.

\bibitem{MS}
F.J. MacWilliams and N.J.A. Sloane, 
\emph{The theory of Error-Correcting Codes,}
The Netherlands, North Holland, 1977.

\bibitem{Par}
K.R. Parthasarathy, 
\emph{An introduction to quantum stochastic calculus,}
Birkh\"auser, 1992.

\bibitem{Spr}
T.A. Springer,
\emph{Invariant theory,}
Lecture notes in mathematics, {\bf 585},
Springer-Verlag, 1977.

\bibitem{Sze}
G. Szeg\"o,
{\sl Orthogonal Polynomials}, 
Colloquium Publications, Vol. 23, New York, AMS, revised eddition 1959,
35--37.

\bibitem{V-J}
D. Vere-Jones,
\emph{Finite bivariate distributions and semi-groups of nonnegative matrices,}
{Q. J. Math. Oxford}, {\bf 22} 2 (1971)
247--270.

\bibitem{YH} R.K. Rao Yarlagadda and J.E. Hershey,
\emph{Hadamard matrix analysis and synthesis: with applications
to communications and signal/image processing,}
Kluwer Academic Publishers, 1997.

\end{thebibliography}
\end{document}